\documentclass[preprint,journal]{vgtc}            

\onlineid{1593}

\ieeedoi{10.1109/TVCG.2025.3633908}

\vgtccategory{Research}

\vgtcpapertype{Technique}

\title{\textcolor{black}{PixelatedScatter: Arbitrary-level Visual Abstraction \\for Large-scale Multiclass Scatterplots}
}

\author{Ziheng Guo, Tianxiang Wei, Zeyu Li, Lianghao Zhang, Sisi Li, 
and Jiawan Zhang*, \textit{Senior Member, IEEE}}
\authorfooter{
\item
 Z. Guo, T. Wei, L. Zhang, S. Li, and J. Zhang(Corresponding Author) are with the College of Intelligence and Computing, Tianjin University, Tianjin 300350, China. E-mail: \{sygzh6, tianxiangwei, lianghaozhang\}@tju.edu.cn; sisilee144144@gmail.com; jwzhang@tju.edu.cn.
\item
 Z. Li is with the State Key Laboratory of Media Convergence and Communication, Communication University of China, Beijing 100024, China. E-mail: lzyfun@cuc.edu.cn.
}

\abstract{%
Overdraw is inevitable in large-scale scatterplots. Current scatterplot abstraction methods lose features in medium-to-low density regions. \textcolor{black}{We propose a visual abstraction method designed to provide better feature preservation across arbitrary abstraction levels for large-scale scatterplots, particularly in medium-to-low density regions.} The method consists of three closely interconnected steps: first, we partition the scatterplot into iso-density regions and equalize visual density; then, we allocate pixels for different classes within each region; finally, we reconstruct the data distribution based on pixels. \textcolor{black}{User studies, quantitative and qualitative evaluations demonstrate that, compared to previous methods, our approach better preserves features and exhibits a special advantage when handling ultra-high dynamic range data distributions.}
}


\teaser{
  \centering
  \includegraphics[width=.95\linewidth]{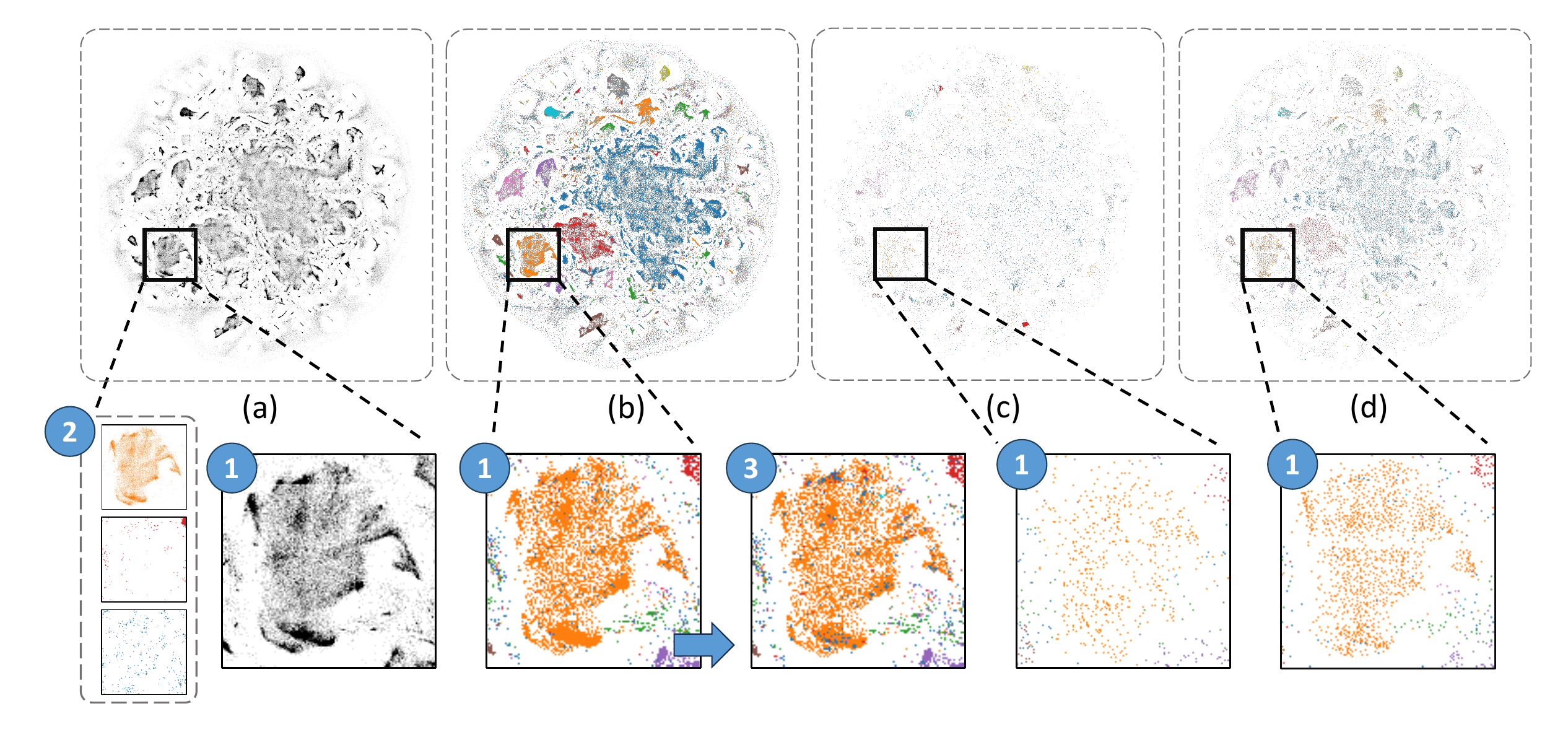}
  \caption{%
  	The representations of different abstraction methods on a scatterplot containing 559,509 points and 10 classes. All images are rendered on an 800$\times$800 canvas. (a) is the density greyscale map of the original scatterplot, where we manually adjusted the density-to-greyscale scale (density ranging from 0 to 20 points per pixel is linearly mapped to greyscale values from 0 to 1) to clearly distinguish low-density regions. (b), (c), and (d) represent our method, MDM (Voronoi + dot density) \cite{Jo_Vernier_Dragicevic_Fekete_2019}, and Recursive Subdivision Based Sampling \cite{Chen_Ge_Zhang_Chen_Fu_Deussen_Wang_2020}, respectively. The results demonstrate that our method better preserves local details in medium-to-low density regions, such as texture, shape, and inter-class relative density (see (b), \Circled{1}, compared to (c) and (d), \Circled{1}). Our method also more effectively maintains density differences between regions (see global regions in (b), (c), and (d)). Furthermore, our method enhances outlier representation while preserving relative class density (see (b), \Circled{3}, with reference to the class distributions in (a), \Circled{2}).%
  }
  \label{fig:teaser}
}

\graphicspath{{figs/}{figures/}{pictures/}{images/}{./}} 

\usepackage{tabu}                      
\usepackage{booktabs}                  
\usepackage{lipsum}                    
\usepackage{mwe}                       

\usepackage{mathptmx}                  
\usepackage[linesnumbered,ruled]{algorithm2e}

\SetCommentSty{mycommfont}

\DontPrintSemicolon
\usepackage{color}
\SetKwRepeat{Do}{do}{while}%
\usepackage{amsmath,amssymb}
\usepackage{enumitem}

\usepackage{amsmath,amssymb}

\usepackage[dvipsnames,svgnames]{xcolor}
\usepackage{circledsteps}        
\pgfkeys{/csteps/inner ysep=3pt}
\pgfkeys{/csteps/inner xsep=3pt}

\begin{document}


\firstsection{Introduction}

\maketitle

Large-scale multiclass scatterplots often suffer from overdraw, where overlapping points in \textcolor{black}{high} density regions obscure the data distribution, leading to the loss of features such as data density and outliers. This severely interferes with various scatterplot tasks such as browsing and exploring, cluster rationalization, and density judgments\cite{Sarikaya_Gleicher_2018}.

\textcolor{black}{In recent years, researchers have proposed numerous methods to mitigate overdraw. Among these, visual abstraction methods have been widely adopted for large-scale scatterplot visualization because they clearly represent data features and offer rendering times independent of the dataset size. Such methods employ a simplified set of visual objects to convey regional features. Some of these methods are based on density estimation and use regional shapes for abstraction, such as continuous density maps\cite{Bachthaler_Weiskopf_2008}\cite{Feng_Kwock_Yueh_Lee_Taylor_2010}\cite{Jo_Vernier_Dragicevic_Fekete_2019} or contours\cite{Collins_Penn_Carpendale_2009}\cite{Li_Zhang_Jia_Zhang_2019}\cite{Tory_Sprague_Wu_So_Munzner_2007}. However, these approaches can introduce visual chaos or synthesize new colors, and by obscuring the intra-region data distribution, they fail to provide a sufficiently detailed representation. Other methods rely on spatial partitioning, dividing the entire space into a series of patches, such as hexbins\cite{Li_Griffith_Becker_2016}\cite{84059e64-8132-3530-83f7-847a77216710} or grids\cite{Luboschik_Radloff_Schumann_2010}\cite{hagh2007weaving}\cite{Jo_Vernier_Dragicevic_Fekete_2019}, and express features on a per-patch basis. However, the data density in most large-scale multiclass scatterplots exhibits a high dynamic range (HDR), meaning there is a vast difference in density values between the densest and sparsest regions. When preserving global density differences, existing methods severely compress the visual representation of vast medium-to-low density regions. This results in significant detail loss, manifesting as large blank or sparse regions. In \cref{fig:teaser}(c), only a few regions are completely filled, while the vast majority of the remaining region, such as the region marked \Circled{1} in (c), appears extremely sparse, losing the relative density (compare with \Circled{1} in (a)) and blue and red outliers (compare with \Circled{2} in (a)). Although \cref{fig:teaser}(d) preserves the outliers, the regional density appears rather uniform, making it difficult to distinguish relatively dense regions from sparser ones.}

\textcolor{black}{To address above issue posed by HDR data, it is necessary to dynamically map data density to visual density, stretching the visual representation of medium-to-low density regions to make their internal details visible. However, the regional density within a patch is often inaccurate. For example, when a patch overlaps with both the dense edge of a cluster and an adjacent empty region, the average density will be a misleading 'medium' value that represents neither the dense nor the empty part accurately. Any visual enhancement based on this flawed regional density will inevitably distort the actual data density. Therefore, a prerequisite is to partition the space into what we call "iso-density regions"—spatially continuous areas of similar density—to ensure the average density is truly representative. Effectively identifying these regions in large-scale scatterplots poses three challenges. First, the regions should be of arbitrary shape to conform to the original data distribution contours. Second, their areas must be dynamically scalable. Finally, the adaptive partitioning process must be efficient for large-scale data. Grid partitioning with a fixed size fails to meet the second criterion, while previous adaptive methods often struggle with the first, typically producing regularly shaped partitions.}

\textcolor{black}{In this paper, we introduce PixelatedScatter, a visual abstraction method for large-scale multiclass scatterplots. Our method addresses all three challenges with a novel partitioning algorithm. It starts with a fast, grid-based partitioning (addressing the third challenge) and yields initial coarse partitions with arbitrary shapes (addressing the first challenge). Then, it adaptively iterates by reducing the grid cell size in denser areas to achieve finer partitions, thus satisfying the second challenge of scalable region sizes. Once this partitioning is established, we employ a density distribution equalization algorithm to adjust visual densities across regions.} To preserve multiclass features based on the regional visual density, we then identify semantic outliers (points not belonging to the same class as their neighbors \cite{9226404}) within each region and balance the preservation of relative class densities with the representation of outliers. This process determines the final pixel number for each class within each region. Finally, we place pixels as closely as possible to the original data distribution, and spread them out with minimum displacement to reconstruct the local data distribution. The final representation of pixels or pixel-like colored squares, which we term a pixelated representation, is a natural representation format for our grid-based approach. Moreover, since our partitioning process depends solely on the data distribution and not on the abstraction level (i.e., the representation level of visual units, where a higher resolution corresponds to a lower abstraction level), our method is inherently suitable for arbitrary levels of abstraction. We demonstrate the effectiveness of our method through a quantitative evaluation involving four metrics compared with related approaches, a user study, as well as two case studies. In summary, the contributions of our work are as follows:

\begin{itemize}
  \item We propose a visual abstraction method for multiclass scatterplots that effectively preserves features across a high dynamic range of data densities, particularly suitable for large-scale datasets.
  \item We propose an efficient and accurate partitioning algorithm that rapidly partitions a scatterplot into arbitrarily shaped iso-density regions.

\end{itemize}

\section{Related Work}

\begin{figure*}[tbp]
    \centering
    \includegraphics[width=\textwidth]{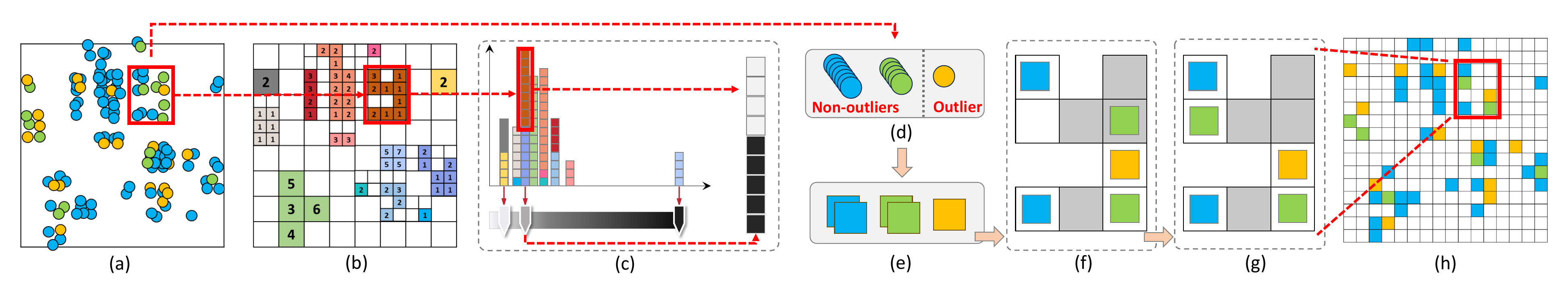}
    \caption{Pipeline of our methods: (a) Original scatterplot; (b) Clusters after iso-density region partitioning, where each color represents a distinct cluster; (c) Splitting each cluster into density-consistent pixels to generate a data density distribution histogram, then equalizing the histogram and using the cumulative distribution function as the mapping to visual density, finally calculating the number of pixels \textcolor{black}{that} need \textcolor{black}{coloring} within each cluster; (d) Filtering outlier classes based on the class distribution within individual clusters; (e) Allocating the pixel number from (c) to obtain class pixel numbers within a cluster; (f) Constructing an initial pixel layout; (g) Dispersing pixels to produce a non-overlapping layout; (h) Final representation result.}
    \label{fig:pipeline}
    \vspace{-.2in}
\end{figure*}

Researchers have proposed various strategies for visualizing scatterplots to mitigate overdraw, which can be broadly categorized into abstraction, appearance adjustment and point mapping.

\subsection{Abstraction Methods}

Abstraction methods typically simplify and optimize the visual representation of data via extracting features and re-encoding the abstracted features. These methods can be further divided \textcolor{black}{into} data space abstraction and visual space abstraction.

\textbf{Data abstraction} \textcolor{black}{alleviates} overdraw by reducing the number of \textcolor{black}{points}. As the reduction in \textcolor{black}{the} quantity of points usually leads to information loss, these methods generally do not aim to preserve all features but design sampling strategies for specific tasks to provide biased but \textcolor{black}{clearer} results. Typically, these methods focus on different goals such as relative densities preservation\cite{Dix_Ellis_2002}\cite{Bertini_Santucci_2006}\cite{Chen_Chen_Mei_Liu_Zhou_Chen_Gu_Ma_2014}\cite{chen2021pyramid}, outlier preservation\cite{Liu_Jian-nan_Liu_Wang_Wu_Zhu_2018}\cite{Xiang_Ye_Xia_Wu_Chen_Liu_2019}\cite{chen2021pyramid} and intra-region relation preservation. Methods biased on density is optimized to preserve the relative density of different regions. However, this focus on density might cause outlier loss. Conversely, outlier biased samplings excel at preserving outliers but may not accurately represent the overall density patterns. Wang et al.\cite{Chen_Ge_Zhang_Chen_Fu_Deussen_Wang_2020} introduced a non-uniform recursive sampling technique for multiclass scatterplots, aiming to preserve major outliers and relative class densities. This KD-tree-based partition, however, can introduce a bias by failing to preserve the spatial arrangement of points within a local region, which may distort the perceived shape of data clusters\cite{9226404}. These methods inevitably introduce task-specific bias and cannot fully eliminate overdraw. It is noted that although work of Wang et al. is a sampling technique, it can be conceptually viewed as a form of visual abstraction. This is because their approach first computes colored grids for each leaf node to represent regional features, and only then selects a point of the same class from the grid as the sample. 

\textbf{Visual abstraction} employs regional shapes to represent areal features, and usually shows clear results in preserving specific features, such as density maps\cite{Bachthaler_Weiskopf_2008}\cite{Feng_Kwock_Yueh_Lee_Taylor_2010}\cite{Jo_Vernier_Dragicevic_Fekete_2019} and contour lines\cite{Collins_Penn_Carpendale_2009}\cite{Li_Zhang_Jia_Zhang_2019}\cite{Tory_Sprague_Wu_So_Munzner_2007}. Although this achieves a more explicit expression of density features and contours, outliers and sparse areas are often overlooked. Some methods incorporate outlier detection and explicitly highlight outliers\cite{feng2010matching}\cite{Mayorga_Gleicher_2013}, while others adaptively segment densities to identify sparse areas\cite{Novotny_Hauser_2006}. 
In these high-level abstraction methods above, color blending and complex visual elements \textcolor{black}{disrupt} the perception of classes, making these methods unsuitable for multiclass scatterplots and incapable of revealing detailed structures.

In contrast to high-level abstraction, low-level abstraction methods assign a patch to each region, typically utilizing color distributions to reflect features. For example, hexbins\cite{Li_Griffith_Becker_2016} is used to preserve the regional density and class density. Multiclass Density Maps\cite{Jo_Vernier_Dragicevic_Fekete_2019} render various mutliclass density map representations based on 2D histograms. 

\subsection{Non-Abstraction Methods}

In contrast to abstraction, non-abstraction methods reduce overlap by optimizing the appearance or position of \textcolor{black}{points}. These methods maintain the correspondence between data \textcolor{black}{points} and visual elements, but they cannot avoid the loss of information that accompanies the rendering of a large number of points on a canvas of limited size.

\textbf{Appearance adjustment} is a straightforward strategy to alleviate overdraw, such as point size reducing\cite{Woodruff_Landay_Stonebraker_1998}\cite{Li_Martens_van_Wijk_2010}, opacity decreasing\cite{Wilkinson_1999} and shape alternative\cite{krzywinski2013plotting}\cite{dang2010stacking}. Recent methods have automated or semi-automated these adjustments based on perceptual theories to enhance scatterplot visual quality. For multiclass scatterplots, many studies focus on color optimization to emphasize relationships across different classes\cite{lee2012perceptually}\cite{wang2018optimizing}. Some research proposes alternative plotting symbols to reduce overlap and enhance perceptual separation, as demonstrated by radial line symbols like the plus sign and the asterisk\cite{tremmel1995visual}. However, complex visual elements introduce new visual biases and significantly increase visual complexity. Since appearance properties lack the scalability to accommodate information (e.g., stacking few semi-transparent points may lead to a opaque area), these methods are typically suitable for small-scale scatterplots without significant overdraw issues.

\textbf{Point mapping} reduces overdraw by dispersing \textcolor{black}{points} while maintaining a one-to-one correspondence between data points and visual \textcolor{black}{units}. Ideally, they excel in feature preservation, especially local structural features while screen resolution is sufficient to display discrete points.  For instance, subspace mapping methods use equidistant grids (Dgrid\cite{hilasaca2019overlap}, IsoMatch\cite{fried2015isomatch}), space-filling curves (HaGrid\cite{cutura2021hagrid}), etc., to create mutually exclusive subspaces, then map data points into these subspaces to eliminate overlap. Point displacement methods\cite{misue1995layout}\cite{hayashi1998layout}\cite{gansner2010efficient}\cite{nachmanson2016node}\cite{li2022dual} spread \textcolor{black}{points} of a certain size from their original positions, generally aiming to minimize dispersing area. However, \textcolor{black}{there is an} inherent conflict between point visibility (generally ensured by point radius) and displacement distortion, making point-to-point mapping methods prone to pixel overdraw.

The Point-to-Pixel Mapping approach\cite{raidou2019relaxing} shifts the visual unit from points to pixels, aiming to better utilize screen space. For example, Gridfit\cite{keim1998gridfit} relocates unallocated data points along predefined curves to the nearest unoccupied pixels to fill space. This method achieves a balance between basic visibility requirements (avoiding pixel overdraw) and reducing distortion. However, displacement inevitably leads to the loss of local structures, and extensive filling can severely impair density and shape features.

\section{Methods}

Our goal is to design a visual abstraction that enhances the visual representation of medium-to-low density regions while preserving the main features of multiclass scatterplots\cite{Chen_Ge_Zhang_Chen_Fu_Deussen_Wang_2020}\cite{Sarikaya_Gleicher_2018} (relative data density, relative class density and outliers). We address this problem using three coherent steps: first, adaptively partitioning the scatterplot and nonlinearly mapping regional data density to visual density to enhance the visual density of medium-to-low density regions under the constraint of preserving relative data density; then, identifying outliers within regions to guide the preservation of relative class density and outlier class; finally, simulating the original data distribution by pixels. \cref{fig:pipeline} illustrates the pipeline of our work.

\subsection{Iso-density Region-based Density Equalization}

Iso-density regions are a series of regions with internally similar density distributions that reflect different data patterns, as changes in underlying data patterns often lead to abrupt density shifts. We utilize this partitioning for the precise estimation of regional data density (see \cref{fig:pipeline}(b)). Then, in order to adapt to arbitrary data distributions, we propose to nonlinearly and monotonically map data density to equalized visual density, which is called regional density distribution equalization\textcolor{black}{(see \cref{fig:pipeline}(d)).} This is inspired by histogram equalization\cite{gonzales1987digital}, as it can monotonically expand the distribution range of data, aligning with our goal. The fundamental basis of the above process is that \textcolor{black}{an} iso-density region almost has no multiple data \textcolor{black}{patterns;} as long as the mapping is monotonic, it preserves the relative density order between regions.


\textbf{Iso-density Region Partition. }The partition has three key characteristics: similar data densities within each region, continuous and irregular shapes of regions, and the ultra-fine granularity of the partitioning, which requires high time efficiency. To meet these characteristics, we employ a grid-based approach combined with a simple clustering strategy to rapidly produce a coarse partitioning result, followed by iterative refinement through sub-partitioning, where sub-partitioning condition depends on the kurtosis of regional density distribution. The method is separated into three parts: gridding, clustering and iterative refinement.

Gridding partitions a scatterplot into square grids with a grid size. Due to the method involving multiple partitioning steps, the grid size varies with the granularity of partitioning. However, for any given scatterplot, we always begin with an initial gridding using a grid size of 1 pixel. This ensures that the result can always be precisely represented by canvas pixels, and is more conducive to understanding. In fact, the initial grid size represents the abstraction scale. In practice, we recommend adjusting the abstraction scale through canvas size; for example, setting the grid size to 2 pixels is equivalent to halving the canvas size, which is both safer and more flexible.

Clustering on grids involves two steps: firstly we count all grids containing \textcolor{black}{points}, referred \textcolor{black}{to} as non-empty grids, where each represents a cluster. Then, we traverse these non-empty grids, merging clusters within a 3 $\times$ 3 grid area centered on each non-empty grid. This guarantees each point in a cluster has another point within a distance of $2\sqrt{2} \times GS$, where $GS$ represents the grid size.

\begin{figure}[tbp]
    \centering
    \includegraphics[width=\columnwidth]{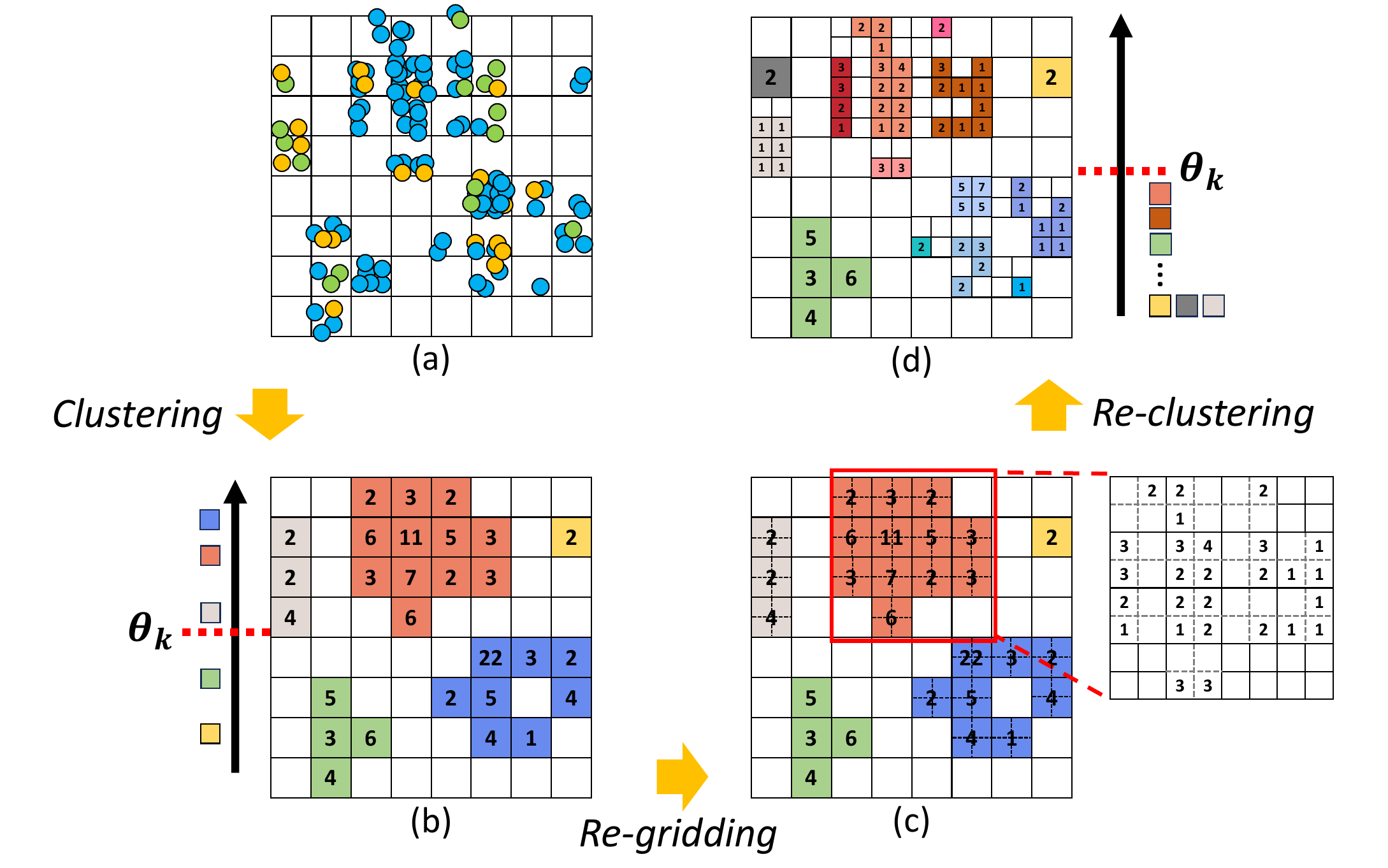}
    \caption{Illustration of Iso-density Region Partition: (a) Gridding of the original data; (b) Clusters after initial clustering, with the kurtosis calculated for each cluster; (c) Re-gridding clusters with kurtosis larger than $\theta_{k}$ using a halved grid size; (d) The Final cluster partition result after re-clustering, where the kurtosis of each cluster is less than $\theta_{k}$.}
    \label{fig:fig3}
    \vspace{-.2in}
\end{figure}

Iterative refinement connects the results of gridding and clustering. The key idea is to sub-partition clusters adaptively through kurtosis. For every cluster, we count point numbers of all grids to get density distribution. Although the distribution is two-dimensional, we are not concerned with locations of valleys and peaks but rather the flatness of the distribution. So we directly calculate the one-dimensional kurtosis of distribution to estimate whether this cluster is uniform. If the kurtosis exceeds a fixed threshold \( \theta_{k} \), we half the grid size, then grid and cluster again, shown in (\cref{algorithm1}). In our algorithm, we use a partitioning level $L$ (relative to the initial gridding) to uniformly adjust multiple clusters to a specified level. Typically, $L$ increases sequentially, representing iterative refinement layer by layer. However, $L$ can also decrease, indicating a re-gridding and clustering with a larger grid size. This iterative refinement process continues until every cluster kurtosis falls below \( \theta_{k} \), as shown in \cref{fig:fig3}(d).

The whole partitioning is as follows: for a given scatterplot, we first perform initial gridding. Subsequently, we use an initial partitioning level $L_{init}$ (default set to -1) to aggregate grids, which \textcolor{black}{help} preserve outliers in sparse region. Then, we apply iterative refinement to obtain the final iso-density region clusters.

\begin{algorithm}[htb]
  \SetKwInOut{Input}{Input}
  \SetKwInOut{Output}{Output}
  \let\oldnl\nl
  \newcommand{\nonl}{\renewcommand{\nl}{\let\nl\oldnl}}

  \Input{\\
      \Indm $DS : {\{ (x, \ y)\} _N}$      \tcp*[f]{Original 2D data}\\ 
      \nonl $\theta_{k}$ \tcp*[f]{Kurtosis threshold}\\
      \nonl $L_{init}$ \tcp*[f]{Initial partition level}\\
      }
  
  \Output{\\                                            
          \Indm $clusters_g: {\{{\{ (x, \ y)\} _{n_i}}\}} _M , ~\sum n_i = N$ \tcp*[f]{Equal-density clusters} \\
  }
  
  \BlankLine
  \SetKwProg{Fn}{function}{}{end~function}
  \SetKwFor{For}{for}{do}{end~for}
  \SetKwFor{While}{while}{do}{end~while}
  \SetKwIF{If}{ElseIf}{Else}{if}{then}{else if}{else}{end~if}

    \BlankLine
    \Fn{\textnormal{Partition(} $DS$, $\theta_{k}$, $L_{init}$ \textnormal{)}}{
        $clusters_{p} \gets {\{ DS \}}$ \tcp*[f]{Clusters peaked need partition}\\
        $clusters_{g} \gets \emptyset$ \tcp*[f]{Clusters gentle that is iso-density}\\
        $L \gets L_{init}$ \tcp*[f]{Partition level}\\ 
        \While{$clusters_{p}$ \textnormal{not} Null}{
            $clusters'\gets \textnormal{SinglePartition(} clusters_{p}, L \textnormal{)}$ \;
            $clusters_{p} \gets \textnormal{FilterKurtosis}(\mathit{clusters'}, \theta_{k})$ \;
            $clusters_{g} \gets clusters_{g} \cup ( clusters' - clusters_{p} )$ \\
            $L$++ \\
        }

        \Return{$ clusters_{g} $}\\
    }

  \BlankLine
  \Fn{\textnormal{SinglePartition(}\textit{clusters}, $L$\textnormal{)}}{
    $\mathit{clusters'} \gets \emptyset$ \;
    \For{\textnormal{\textit{cluster} in \textit{clusters}}}{
        $D \gets \textnormal{Gridding(}cluster, \frac{1}{2^{L}}\textnormal{)}$ \tcp*[f]{Re-gridding with half grid size}\;
        $clusters' \gets clusters' \cup ~ \textnormal{Clustering(}D \textnormal{)}$
    }
    \Return{$clusters'$ } \;
  }

  \caption{Iso-density Region Partition}
  \label{algorithm1}
\end{algorithm}

\textbf{Regional Density Distribution Equalization. } 
First, for each iso-density region (hereafter "cluster"), we calculate its data density by dividing the number of points it contains by its area in pixels. Since these cluster regions can overlap, a single pixel location on the canvas might be associated with multiple clusters. We resolve such conflicts by assigning the pixel to the cluster with the smallest area-to-class-number ratio, a heuristic that prioritizes the preservation of more complex, multiclass features.

\begin{figure}[tbp]
    \centering
    \includegraphics[width=\columnwidth]{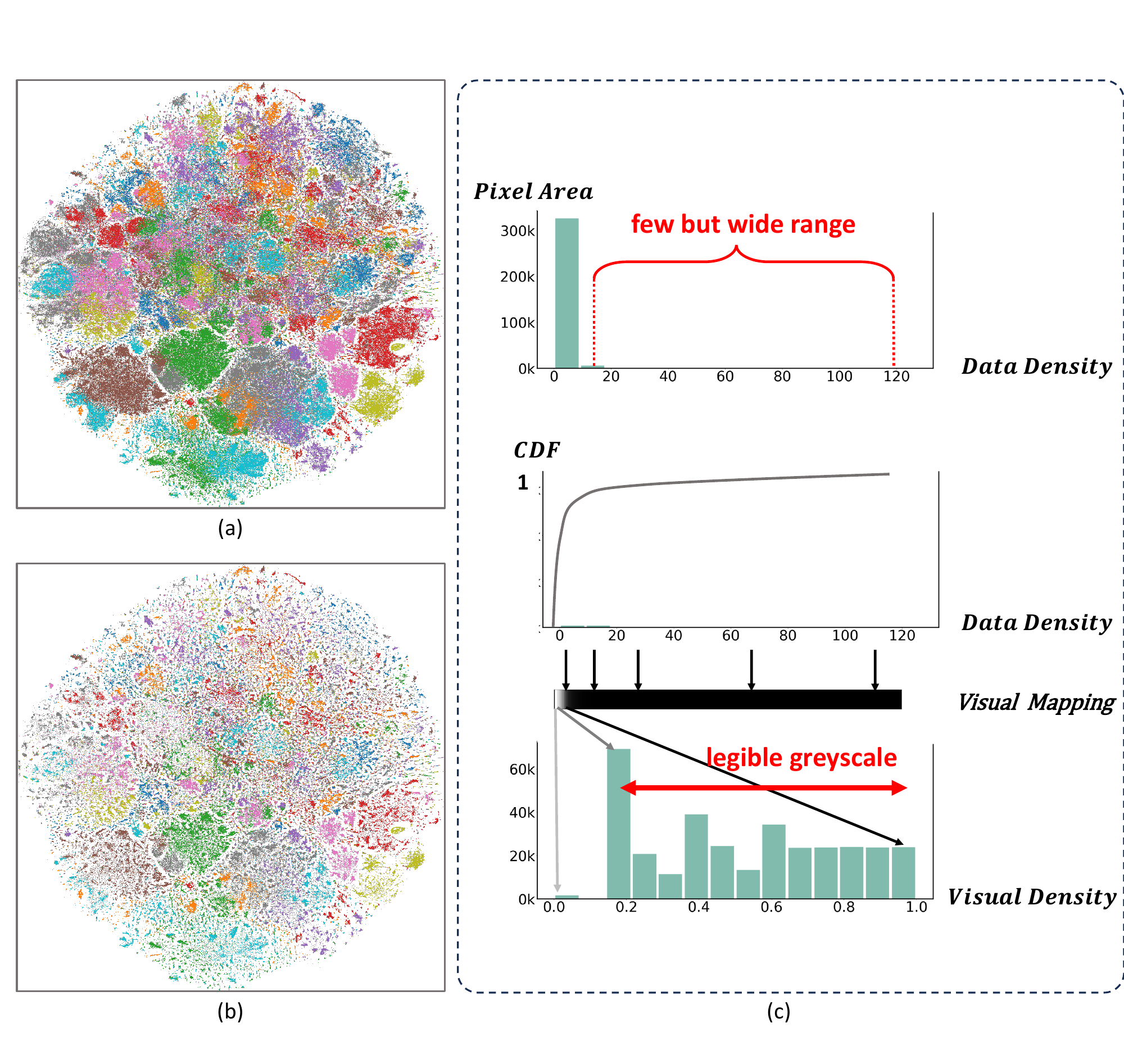}
    \caption{Regional Density Distribution Equalization: (a)  the original scatterplot; \textcolor{black}{(b) effect} after density equalization; (c) density equalization mapping process, where the top section shows the original data distribution, the middle section displays the cumulative distribution function after equalization, and the bottom section presents the distribution of visual density after mapping.}
    \label{fig:density grey map}
    \vspace{-.2in}
\end{figure}

With the data density for each region established, we then construct a data density histogram. This histogram is built by considering the set of all regions; specifically, for each region, its area (in pixels) is added as a count to the histogram bin corresponding to that region's data density value. Therefore, the total population for the equalization is the sum of the areas of all regions. Next, we compute the Cumulative Distribution Function (CDF) from this histogram. The core of our mapping is this: the new visual density for a region is defined as its corresponding value in the CDF. Because the CDF is normalized between 0 and 1, this effectively assigns visual densities "as probabilities," stretching the density range to enhance contrast, as depicted by the greyscale mapping bar. Finally, the number of points to be rendered in each region is recalculated based on this new, equalized visual density. The resulting layout and its more uniform visual density histogram are shown in \cref{fig:density grey map}(b) and at the bottom of \cref{fig:density grey map}(c), respectively.

\subsection{Outlier-guided Class Pixel Allocation}

After partitioning a scatterplot into iso-density region clusters, a single cluster may contain points of different classes. If we allocate pixels faithfully according to class proportions, outlier classes may not be preserved. Conversely, if we excessively enhance the representation of outlier classes, the relative class density may be compromised. Thus we defined the characteristic differences between outliers and non-outliers and proposed an outlier class filtering model. Based on this, two stages of formalized expression for class outlier emphasis are introduced to address the conflict between the outlier representation and relative class density preservation.


\begin{figure*}[tbp]
    \centering
    \includegraphics[width=.85\textwidth]{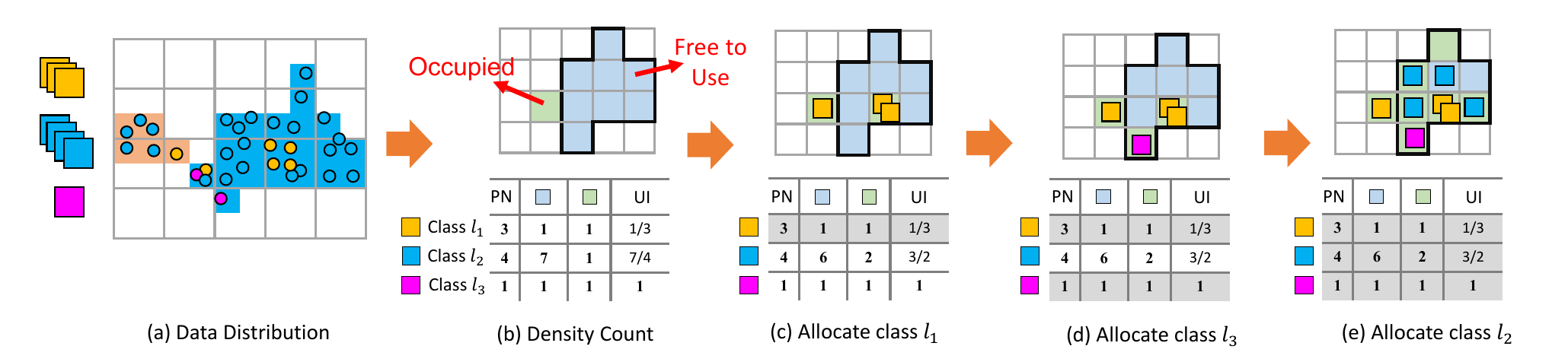}
    \caption{Pipeline of Initial Layout Construction: (a) The initial state; (b) Identifying placeable (light blue) and unplaceable (light green) regions; (c, d, e) Iteratively placing the most urgent class after calculating urgent indexes (UI) to complete the initial layout.}
    \label{fig:fig4}
    \vspace{-.2in}
\end{figure*}


\begin{figure*}[tbp]
    \centering
    \includegraphics[width=.85\textwidth]{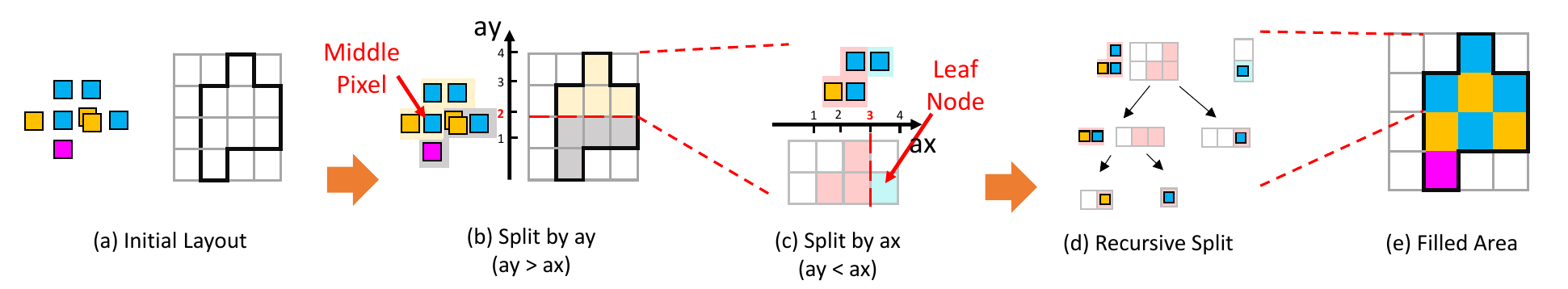}
    \caption{Pipeline of KD-Tree Guided Pixel Dispersion: (a) The initial overlapping layout; (b) Splitting the region along its longer axis (the y-axis); (c, d) depict the sub-split within the yellow area, and such splits continue until each pixel is assigned to just one grid; (e) The non-overlapping pixel layout.}
    \label{fig:fig5}
    \vspace{-.2in}
\end{figure*}

\textbf{Outlier Classes Filtering. }To accurately distinguish between outlier and non-outlier classes, we focus on two aspects: First, ensuring that the point number of each outlier class does not exceed a threshold \textcolor{black}{$\tau_{o}$}$\times N_{avg}$, where $N_{avg}$ represents the average point number of classes. The introduction of $N_{avg}$ \textcolor{black}{is because the proportions of classes are related to the class number; the proportion of each class naturally decreases as the number increases.} Second, the total proportion of non-outlier classes must be no less than a threshold \textcolor{black}{$\tau_{ns}$}. These two thresholds ensure the minority of outliers while maintaining differences among non-outliers. These two thresholds are related, specifically:

For a $n$-classes cluster containing $N_c$ points, the average point number of classes $N_{avg}$ is $\frac{N_c}{n}$. The total number of outlier classes must not exceed $(n-1)\textcolor{black}{\tau_{o}} \frac{N_c}{n}$, reflecting an extreme case where the cluster contains only one non-outlier class. Consequently, the total number of non-outlier classes is no less than $(1 - \frac{n-1}{n}\textcolor{black}{\tau_{o}}) N_{c}$, which simplifies to the total proportion of non-outlier classes no less than $1 - \frac{n-1}{n}\textcolor{black}{\tau_{o}}$. This indicates that for a given total proportion of non-outlier classes threshold \textcolor{black}{$\tau_{ns}$}, we can always compute a \textcolor{black}{$\tau_{o}$} for any cluster with different classes numbers, ensuring the \textcolor{black}{$\tau_{ns}$} constraint is satisfied:

\begin{equation}
     \textcolor{black}{\tau_{o}} \leq (1 - \textcolor{black}{\tau_{ns}})\frac{n}{n-1},
\end{equation}

where $0.5 \leq \textcolor{black}{\tau_{ns}} \leq 1$, because the area of non-outliers should not be less than that of outliers. In our method, we set the parameter \textcolor{black}{$\tau_{ns}$} to 0.5 by default.

\textbf{Outlier Emphasis Patterns. } Through the aforementioned filtering, we distinguish between outlier and non-outlier classes in any cluster. Then we propose two stages of outlier emphasis to guide the allocation of pixels quantities:

We assume that class $i$ occupies $p_i$ of point numbers in the region, with the total area being $m$ pixels, so the ideal area with complete density preservation can be calculated as $A_i = p_i  m$. The number of outlier class $i$ is referred as $ON_i$.

\textbf{ST1}: We ensure the presence of pixels for semantic outliers via mandating the minimum pixel number more than one,

\begin{equation}
     ON_{i} = \left\{\begin{matrix}
 1 & if \;\; A_i<1\\
 \left [  A_i \right ] \ & else
\end{matrix}\right.  
\end{equation}

\textbf{ST2}: We ensure the \textcolor{black}{non-inversion} between outliers and non-outliers via mandating the area occupied by the most dense outliers not to exceed that of the least dense non-outliers, where h adjusts the degree of outlier emphasis.

\begin{equation}
     ON_{i} = \left\{\begin{matrix}
 1 & if \;\;  hA_i<1\\
 \left [ hA_i \right ] \ & else
\end{matrix}\right.
    \quad (1 < h \leq \textcolor{black}{h_{max}} )
\end{equation}

The density reversal threshold \textcolor{black}{$h_{max}$} can be given by the following equation, where $op$ and $np$ are referred as outliers proportion and non-outliers proportion; $j$, $k$ are class numbers of outliers and non-outliers:
\begin{equation}
    \max{op}  \textcolor{black}{h_{max}} = \frac{\min{np}}{ \sum_{1}^{k}np_i} (1- \textcolor{black}{h_{max}}\sum_1^j{op_i})
\end{equation}

When $h=1$, the outlier pixel numbers corresponds to that in \textbf{ST1}, indicating the retaining of outliers without emphasis. When $h =\textcolor{black}{h_{max}}$, the area occupied by the most dense outlier class equals that by the least dense non-outlier, reaching the threshold that maintains the non-reversal of density between outliers and non-outliers.



In ST1, nearly uniform density proportion preservation was applied to all classes. In ST2, density order preservation among classes was maintained, while relative proportion was preserved separately in outliers and non-outliers. \cref{fig:application -2} shows the results for h values of 2 and 10.

\subsection{Pixel-based Data Distribution Reconstruction}

Our goal is to reconstruct the data distribution by arranging multiclass pixels within cluster regions according to the original data distribution. To achieve this goal, we design a two-stage pixel layout method: First, we build an initial pixel layout based on the original data distribution. Subsequently, we propose an overlap-free layout method based on first stage. \textcolor{black}{We} use a median-split kd-tree to recursively partition the space and disperse overlapping pixels from the initial layout.
Fig. \ref{fig:fig4} and \ref{fig:fig5} illustrate the pipeline of these two steps respectively.

\textbf{Initial Layout Construction.} We first construct an initial overlapping layout, which is strategically designed to minimize displacement costs in stage two by reducing initial overlaps. The core is a heuristic that prioritizes classes with fewer available grids for placement. For instance, a class with many potential locations can be deferred, as it will likely still have sufficient space after other classes are placed, like the blue class in \cref{fig:fig4}(a). We formalize this by calculating an 'urgent index' (UI) for each class (placeable grids / class size). Pixels are then allocated to all placeable grids. If there are still unallocated pixels, they are placed cyclically according to density from highest to lowest. \cref{fig:fig4} illustrates the above process.

\textbf{KD-Tree Guided Pixel Dispersion.} After obtaining the initial pixel layout, we recursively disperse each pixel to a unique grid using a median-split kd-tree\cite{10234687, raidou2019relaxing}, as shown in \cref{fig:fig5}. A kd-tree is a space-partitioning data structure that organizes points in a multi-dimensional space; for our 2D pixel layout, it builds a binary tree by recursively splitting the set of pixels along alternating axes (x and y). Each split includes two key steps: deciding whether to split along the x-axis or y-axis and determining the position of split line. The first step based on the bounding rectangle of the region to be \textcolor{black}{split. If} the rectangle's (x-axis) width is greater than the (y-axis) height, then split along the x-axis; otherwise, split along the y-axis. This step helps maintain the balance of the binary tree. For the second step, we take the median x-coordinate (or y-coordinate) of the pixels as the split line.

\begin{figure*}[tbp]
    \centering
    \includegraphics[width=\textwidth]{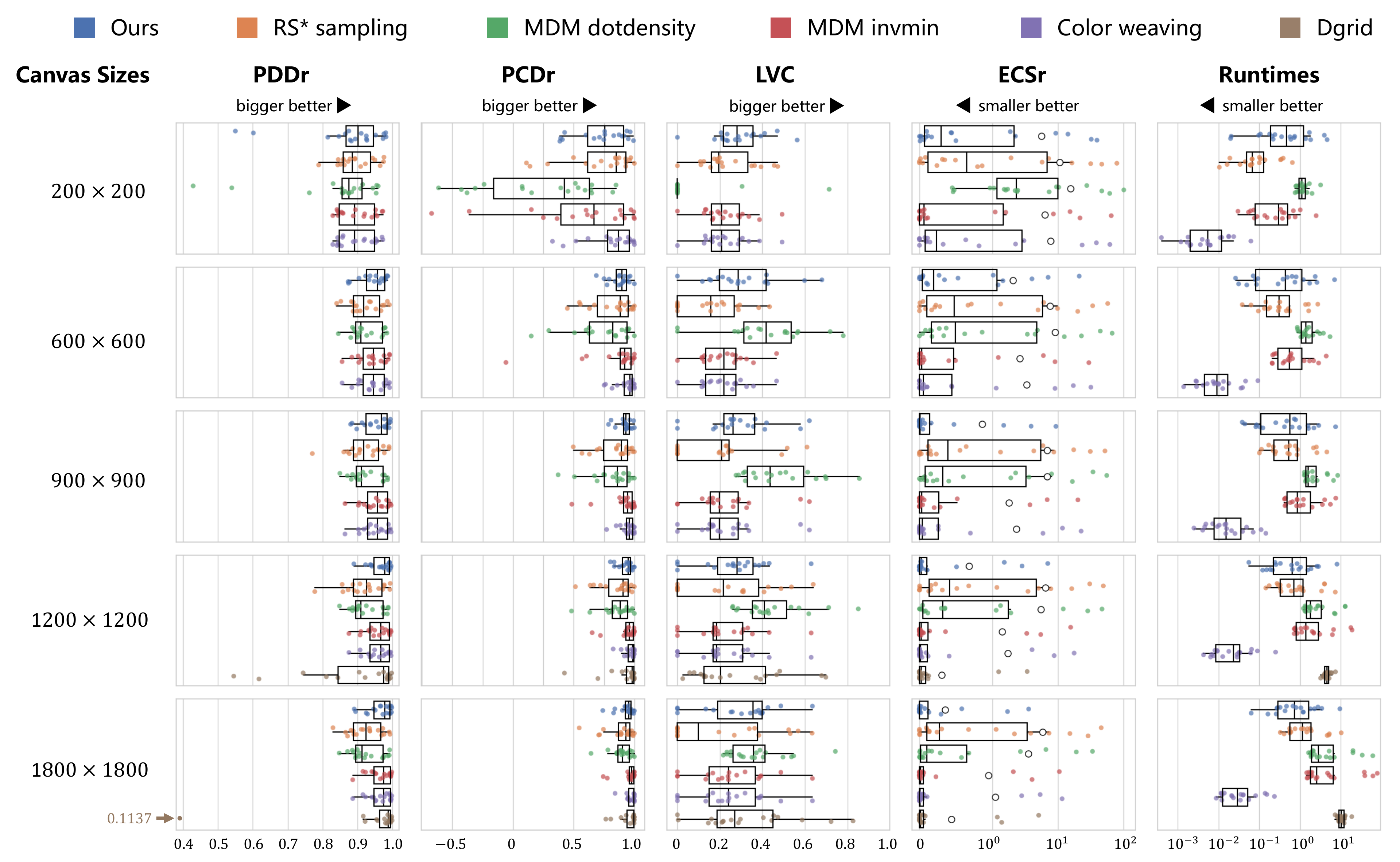}
    \caption{\textcolor{black}{Box plots were used to visualize the results for the four metrics and layout times across 5 canvas sizes.} For PDDr, PCDr, and LVC, higher values indicate better performance; for ECSr, lower values are preferred.}
    \label{fig:evaluation}
    \vspace{-.2in}
\end{figure*}

\section{Evaluation}

\subsection{Quantitative Evaluation}

\textcolor{black}{This section presents the quantitative comparison involving four metrics at different levels of abstraction, qualitative comparison, and a user study conducted to determine whether the results are perceptible to users.} Experiments and rendering were performed on an environment with i9-13900HX 2.2GHz processor, RTX 4060 Laptop GPU, and \textcolor{black}{48GB} RAM. \textcolor{black}{All code was implemented in JavaScript and uesd WebGL for fast browser-side rendering.}

\textbf{Datasets} For evaluating our method, we utilized 20 real-world multiclass datasets from the UCI data repository\cite{Asuncion_2007} and \textcolor{black}{Nomic Atlas data}\cite{nomic_atlas_brandon_collection}, \textcolor{black}{with the number of data points ranging from 13K to 3.6M and the number of classes from 3 to 128. Among them, 11 datasets contain over 400K points, and 4 are at the million scale.}

\textbf{Comparison Methods} We implemented pixel versions of all the following abstract layout methods: Recursive Subdivision based Sampling(bbreviated as RS* Sampling)\cite{Chen_Ge_Zhang_Chen_Fu_Deussen_Wang_2020}, Multiclass Density Maps\cite{Jo_Vernier_Dragicevic_Fekete_2019} and Color Weaving\cite{Luboschik_Radloff_Schumann_2010}. \textcolor{black}{For broader evaluation, the state-of-the-art point mapping method DGrid\cite{10234687} was also included, although such methods are not suitable for arbitrary canvas sizes. Generally speaking, these methods are grid-based; however we set the grid size to a single pixel. The RS* Sampling parameters are set to \( \lambda = 0.02 \) and \( \tau = 0.02 \), following the default settings in the paper. MDM has two sets of parameters: the first uses \(Voronoi\) and \(Dotdensity\) to preserve regional density (with \(rebin_{size}\) set to 2000 and \( rescale \) set to \( sqrt \) to enhance the representation of low-density areas); the second uses \( Invmin \) to highlight outliers. Color Weaving employs random weaving to preserve class density. For DGrid, the gaussian kernel size \( M \) was fixed at 31, as a practical trade-off between performance and acceptable runtime. Our method uses fixed parameters threshold \( \theta_k = 10 \) and visual emphasis ratio \( h = 10 \), while the initial level \( L_{\text{init}} \) was automatically determined as described in the parameter analysis.} 

\textcolor{black}{Although Pixel-Relaxed Scatter Plots\cite{raidou2019relaxing} is also a pixel-based technique, it was primarily designed to optimize the use of the screen space rather than preserve the data distributions of scatterplots. Specifically, the method generates a solidly tiled region that completely disrupts the original distribution’s shape; and its color encoding is designed for single class. These issues render the method unsuitable as a competing approach. Concrete results are provided in supplementary material.}

\textbf{Metrics} Since our method focuses on preserving density information and outliers, the metrics Perceived Data Densities Ratio (PDDr), Perceived Class Densities Ratio (PCDr), and Erased Class Sample ratio (ECSr), summarized by Wang et al.\cite{Chen_Ge_Zhang_Chen_Fu_Deussen_Wang_2020}, align with our objectives and can help evaluation. Additionally, considering human perception of density is nonlinear, especially in large-scale \textcolor{black}{scatterplots, there} are many regions with similar data densities, or extremely high (or low) densities, where the human eye cannot clearly discern the visual density of these regions. Thus, greater visual contrast aids in aligning perception with data density. We introduce a new metric, Legible Visual Contrast (LVC), to measure the clarity of the layout’s visual representation. We provide formal definitions as follows:

Given a pixelated map as \( V: \mathbb{R}^2 \rightarrow \mathbb{R} \), indicating that each grid can be covered by a single label or be empty. A region \( \Omega_i \)'s visual density \( D_v(\Omega_i) \) is defined as the ratio of covered grids and all grid count.

Legible Visual Contrast (\textbf{LVC}) measures the overall visual clarity of a layout, quantified as the average perceivable contrast. Perceivability implies that if the visual density of two regions both does not exceed the visibility visual density threshold \textcolor{black}{$\theta_{grey}$}, then their contribution to the contrast is disregarded. That's because sparse distributed pixel areas are challenging to observe with clear edges, making it visually difficult to distinguish such areas from other similarly sparse regions. Drawing upon well-established principles for calculating contrast in image processing\cite{panetta2013no}, we define the local contrast \(C_{ij}\) between two regions \(\Omega_i, \Omega_j\) as the normalized difference in visual density:

\begin{equation}
\setlength{\abovedisplayskip}{3pt}
    C_{ij} =\left\{\begin{matrix}
0 & if~~ \max({D_v(\Omega_i ), D_v(\Omega_j )}) \le \textcolor{black}{\theta_{grey}} \\
\frac{|D_v(\Omega_i )-\textcolor{black}{\overline{D}_v (\Omega_i ,\Omega_j )}|}{\textcolor{black}{\overline{D}_v (\Omega_i ,\Omega_j )}} & else
\end{matrix}\right.
\setlength{\belowdisplayskip}{3pt} 
\end{equation}
where \(\overline{D}_v(\Omega_i, \Omega_j) = \frac{D_v(\Omega_i) + D_v(\Omega_j)}{2}\) is the average visual density of the two regions. Given this, LVC is then calculated as the mean of local contrasts across all neighboring region pairs:

\begin{equation}
    LVC = \frac{\sum_i\sum_{j~\in neighbors}C_{ij}}{RN}
\end{equation}
where \(RN\) is the total number of region pairs that meet \textcolor{black}{$\theta_{grey}$} condition. The range of LVC is [0,1], where values closer to 1 indicate better overall visual clarity of the layout.

\textbf{Results} 
All abstraction methods were calculated on canvas sizes of 200, 600, 900, 1200, and 1800. Because the canvas size required by DGrid depends on point numbers, it was evaluated only at 1200 (excluding dataset \textit{Tweets from MUSC}) and 1800 (all data). To ensure fair comparison across canvases, the sliding-window size was scaled proportionally with the canvas (to 20, 60, 90, 120, and 180 pixels, respectively), so the analysis window always covered the same relative spatial region. We fix LVC window size at 10 and \(\theta_{grey}\) at 0.3, because of the constancy of the human perceptual. Evaluation results are shown in \cref{fig:evaluation}; for ECSr, where the medians of different methods are close, we additionally annotate the mean with black-edged circles. At almost every canvas size our method consistently leads in both PDDr and ECSr. For PDDr it always attains the highest median with smaller extrema, demonstrating the stability of our region-density preservation strategy. For ECSr most methods score well because many datasets have few classes, masking differences. However, our method achieves the smallest minima and means at every canvas size, even outperforming the outlier-oriented MDM Invmin, proving its comprehensive ability to retain outliers. In PCDr our method ranks second, close to the best. Note that due to our explicit calculation of density and class preservation, there are minor rounding errors in our method, which may cause us to perform slightly worse on sparse, small-scale datasets. However, feature preservation in such datasets is relatively straightforward, and all methods score comparably high. These errors are negligible on large datasets, where our method achieves higher PCDr.


\begin{figure}[tbp]
    \centering
    \includegraphics[width=\columnwidth]{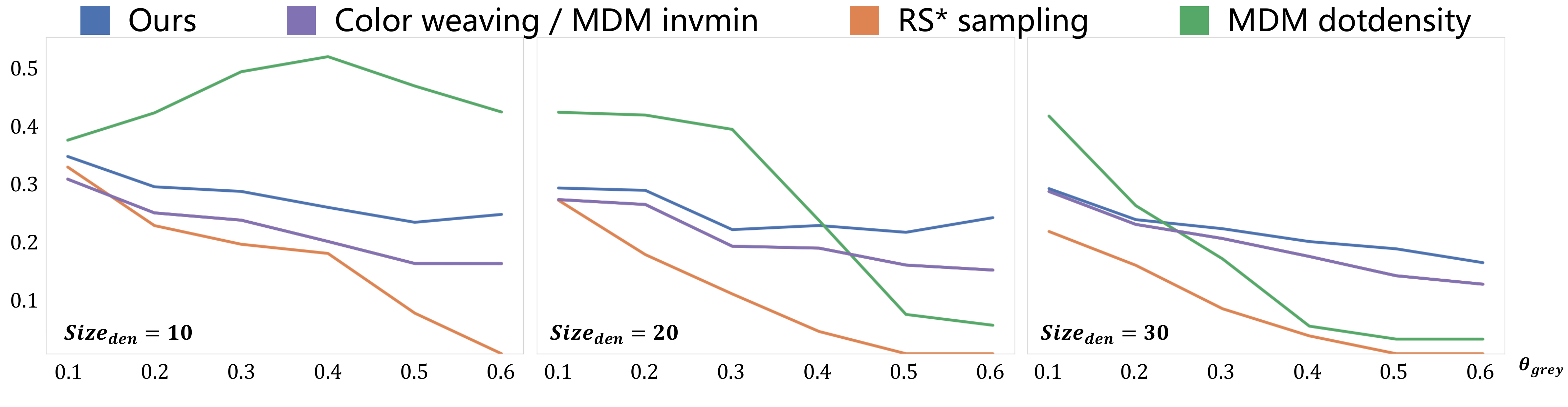}
    \caption{\textcolor{black}{The results of average LVCs on the 900$\times$900 canvas size.} Our method achieves better performance and stability when parameters change. }
    \label{fig:LVC}
    \vspace{-.2in}
\end{figure}

\cref{fig:LVC} shows the LVC trend on a $900\times900$ canvas under different window sizes \( Size_{den} \) and \( \theta_{grey} \). High sensitivity to \( Size_{den} \) suggests that contrast is scale-dependent, while a strong dependence on \( \theta_{grey} \) indicates that the metric is artificially inflated because most regions have very low grey levels. Our method is the most stable and well-performed, indicating the reliable visual density enhancement. RS* Sampling lacks pronounced density differences and maintains similarly low density globally. MDM Dotdensity produces many nearly blank regions and a few dense ones, yielding strong contrast only between them. The contrasts achieved by MDM Invmin and Color Weaving are highly dependent to datasets and largely unrelated to canvas size.

\textcolor{black}{Our method is also the most adaptive to canvas size: as the canvas shrinks its performance decays slowly and stably across all metrics. DGrid requires many empty grids to convey regional information, degrading severely on smaller canvases. Among density-based methods, RS$^{\ast}$ Sampling performs relatively better on small canvases because pixel-based density estimation becomes more uniform. However its metrics improve slowly with larger canvases, with ECSr and PDDr plateauing beyond 600. MDM\,Dotdensity behaves similarly; additionally, its large blank regions yield poor ECSr and PCDr. Other simple strategies (MDM\,Invmin and Color Weaving) deteriorate rapidly on small canvases: MDM\,Invmin over-emphasises outliers on 200 canvas size, sharply reducing PCDr, while Color Weaving’s random sampling causes a marked drop in ECSr at the same size.}

\textcolor{black}{In terms of runtime, although not the fastest, our method processes most datasets within one second and returns results within ten seconds even for million-point, hundred-class datasets. Interestingly, whereas other methods slow down markedly with increasing canvas, ours shows little correlation with canvas size: at 1800 it is already the second fastest. This aligns with our design, which depends on data distributions rather than manual abstraction levels.}

\textbf{Time Complexity} Considering the time complexity of  iso-density region partition: assuming the total number of \textcolor{black}{points} in the dataset is \( N \) and the initial number of grids is \( G \), the overall complexity is \( d  O(N) + d  O(G) \), where \( d \) is the number of partitions. Since \( d \) has an upper bound, making the final time complexity \( O(N) + O(G) \). Considering the time complexity of the subsequent layout process: the outlier separation of all clusters has a time complexity of \( O(G) \). Pixel allocation consists of two steps: the initial pixel assignment with a time complexity of \( O(G) \), and spatial partitioning and remapping using a kd-tree with a time complexity of \( O(G \log G) \). Consequently, the overall time complexity of our method is \( O(G \log G) + O(N) \).

\subsection{User Study}

\textcolor{black}{Although quantitative metrics provide insights, how well PDDr, PCDr, and ECSr reflect pixel-level human perception remains unclear. We therefore designed a user study to directly verify the perception of density and outlier information.}

\textcolor{black}{\textbf{Experimental design.} The study comprised three tasks: \textbf{E1}~relative regional density perception, \textbf{E2}~relative class density perception, and \textbf{E3}~semantic outlier perception. For \textbf{E1} and \textbf{E2} we adopted the protocol of Yuan et al\cite{9226404}. Specifically, \textbf{E1} asked which contained more points for two randomly selected non-empty regions; \textbf{E2} asked which class was more numerous within a given region. For \textbf{E3} we redesigned the task: for each scatterplot two classes were randomly selected; for each class we sampled five regions, each time choosing with equal probability either a random region containing the category or one without it. Participants need to identify which of five regions contained the target class. Effectiveness was measured by accuracy and false-positive rate. Because large scatterplots contain fine structures, the region size was set to $canvas\ size/10$, consistent with the quantitative evaluation, and experiments were conducted at canvas sizes of $600$, $1200$, and $1800$. To maintain perceptual continuity within a given canvas size, we partitioned the three experiments into nine sets by canvas size and randomized question orders within each set to mitigate learning effects. Further details appear in the supplementary material.}

\textcolor{black}{\textbf{Datasets.} We selected seven of the previous twenty datasets: \textit{Forest covertype}, \textit{Hathi trust library}, \textit{Dc census citizens}, \textit{Cup98LRN}, \textit{MoCap}, \textit{CS rankings}, and \textit{Daily sports}. Each of them contains 100 thousands of points, ensuring discernible differences among regions at the region size chosen before. These datasets vary widely: \textit{Cup98LRN} and \textit{CS rankings} exhibit complex inter-class relationships, whereas \textit{Hathi trust library} has regions with diverse densities.}

\begin{figure}[tbp]
    \centering
    \includegraphics[width=\columnwidth]{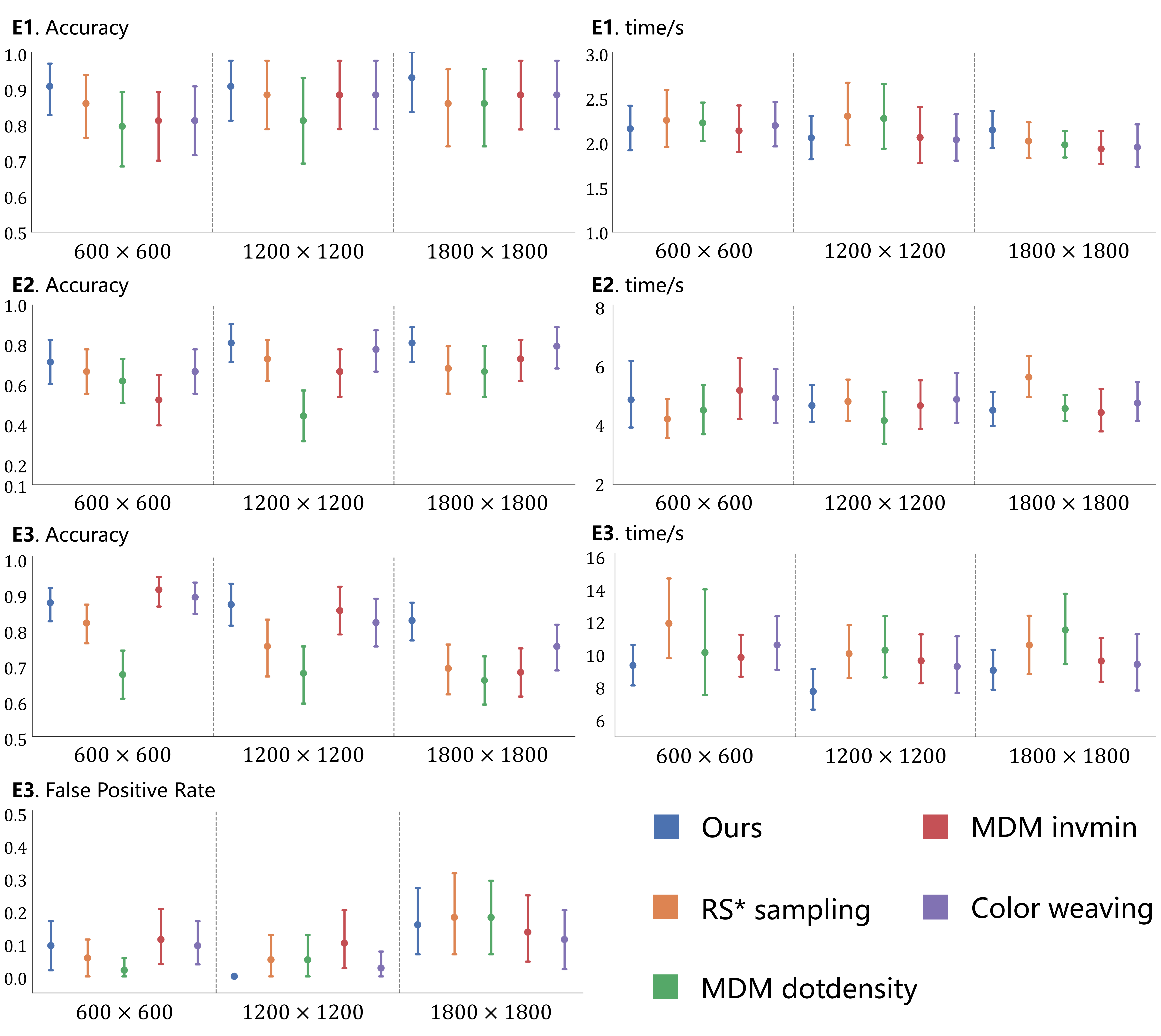}
    \caption{Average accurary and complete time of \textbf{E1}.relative regional density perception and \textbf{E2}.relative class density perception. Average accurary, complete time and False Positive Rate(FPR) of \textbf{E3}.semantic outlier perception.}
    \label{fig:user study}
    \vspace{-.2in}

\end{figure}

\begin{figure*}[tbp]
    \centering
    \includegraphics[width=\textwidth]{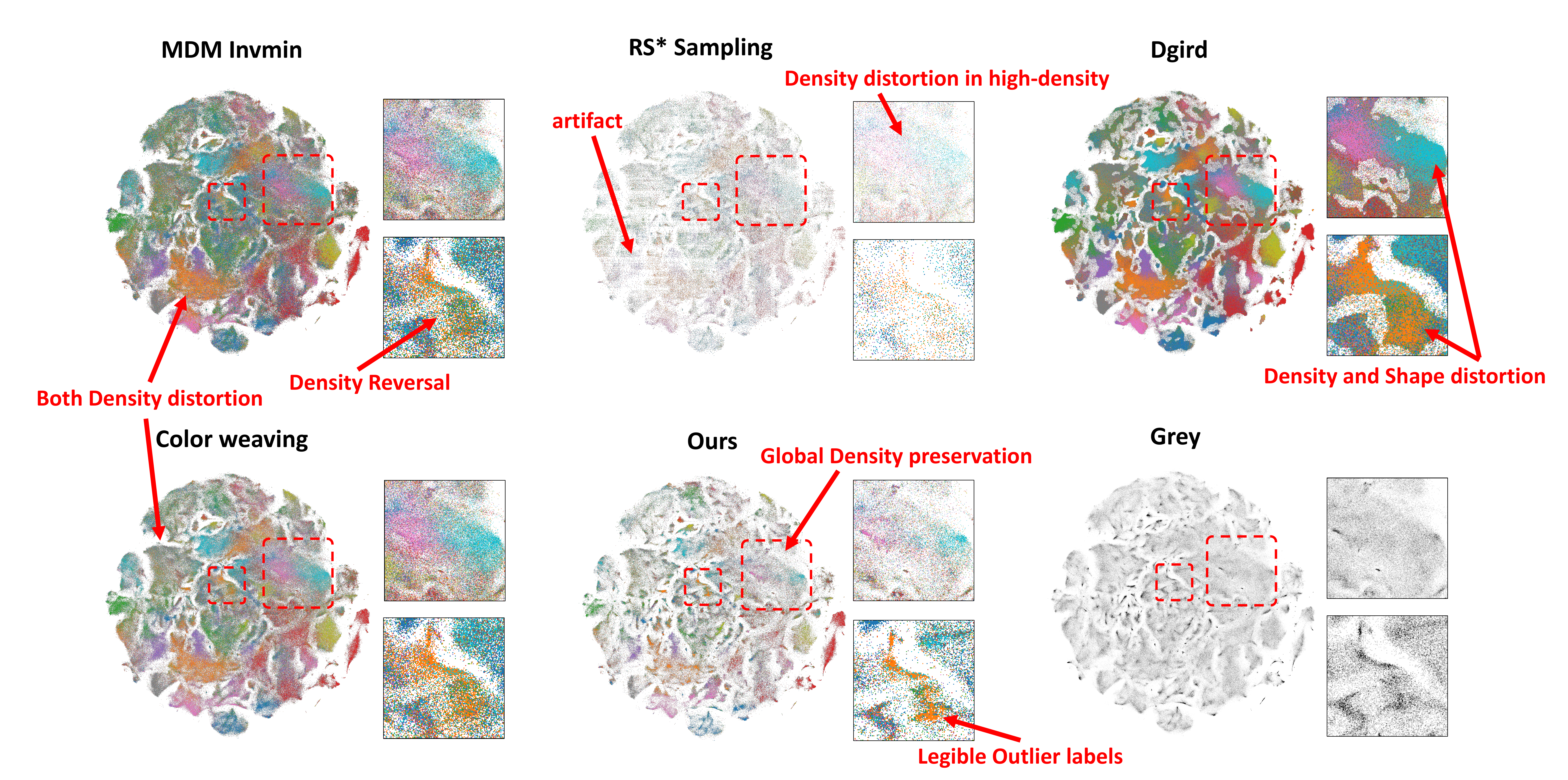}
    \caption{Detailed comparison results on \textit{DBLP samples}. The first five images show results from different methods, followed by a grayscale density map. This map linearly converts density values in the range [0, 7] to grayscale [0, 1], as this range captures over 99.5\% of the data points.}
    \label{fig:Qualitative Evaluation - 1 dataset}
    \vspace{-.2in}
\end{figure*}

\textcolor{black}{\textbf{Participants.} Twenty undergraduate or graduate students were recruited, eight with research experience in visualization or computer graphics. No participant reported color blindness or color weakness.}

\textcolor{black}{\textbf{Apparatus.} The study was conducted offline via a web interface on monitors with a resolution of $1920\times1080$. Before each session, a 20-minute briefing ensured that participants fully understood the background, procedure, and precautions.}

\textbf{Results.} \cref{fig:user study} reports accuracy and completion time for all tasks (\textbf{E3} additionally shows FPR). For relative regional density perception(\textbf{E1}), our method leads at every canvas size, attaining almost \(0.9\) accuracy; MDM Invmin and Color Weaving perform noticeably worse on the small canvas. Friedman tests indicate statistically significant differences in accuracy among these methods \(\) ($p = 0.0362, 0.0625, 0.1070$ for canvas sizes of $600\times600$, $1200\times1200$, and $1800\times1800$ respectively; all subsequent p-values follow this order). For relative class density perception(\textbf{E2}), our method and Color Weaving share the top rank, whereas the outlier-oriented strategy of MDM\,Invmin more readily inverts class density on small canvases \(\) ($p = 0.0252, 0.0153, 0.0112$). Recognition times in \textbf{E1} and \textbf{E2} are short and similar across methods \(\) (for \textbf{E1}, $p = 0.6268, 0.7603,  0.2674$; for \textbf{E2}, $p=0.4830, 0.6151, 0.3386$).

For semantic outlier perception(\textbf{E3}), on the 600 canvas size MDM Invmin and Color Weaving facilitate outlier perception due to more colored pixels; our method performs closely. As canvas size grows, outlier perception declines for all methods because the absolute number of colored pixels per region increases, making outliers harder to spot. Because of the outlier-enhancement strategy, our method amplifies outlier visibility, significantly outperforming others and declining more slowly in terms of accuracy \(\) ($p = 0.0117, 0.0245, 0.0382$). The strategy also yields easily recognizable outlier class clusters, giving our method the shortest completion times on all canvases \(\) ($p = 0.0910, 0.1290, 0.0306$). FPR appears correlated with the confusion of class distributions: our method maintains a comparatively low FPR across all canvas sizes \(\) ($p=0.1875, 0.1598, 0.2081$).

\subsection{Qualitative Evaluation}

To further explore the effectiveness of methods in terms of density and outlier preservation, we conducted a qualitative comparison on the large-scale multiclass dataset \textit{DBLP samples}, which contains 820,000 points across 34 classes, shown in \cref{fig:Qualitative Evaluation - 1 dataset}. All methods were rendered on a 1200 $\times$ 1200 canvas. To facilitate the comparison of global density preservation, we provide the greyscale density map.

Both MDM Invmin and Color Weaving exhibited obvious density distortion because all pixels with points were rendered. This caused the visual density to be related to the area occupied by \textcolor{black}{points} rather than the data density. While MDM Invmin strategy can highlight outliers in most cases, it leads to an imbalance in inter-class density and suffers from density reversal, where low-density but uniformly distributed classes occupy more pixels than the density. In Dgrid, displacement disrupts local structures, with noticeable distortion particularly in high density areas, and large-scale scatterplots require additional canvas space. Our method also allows slight pixel displacement to maintain inter-class density, but our displacement is controlled and does not exceed the original region. Additionally, our nonlinear mapping of visual density enhances contrast for better density perception, and our outlier highlighting strategy safely accentuates outlier representation.

\subsection{Parameter Analysis}

\begin{figure}[tbp]
    \centering
    \includegraphics[width=\columnwidth]{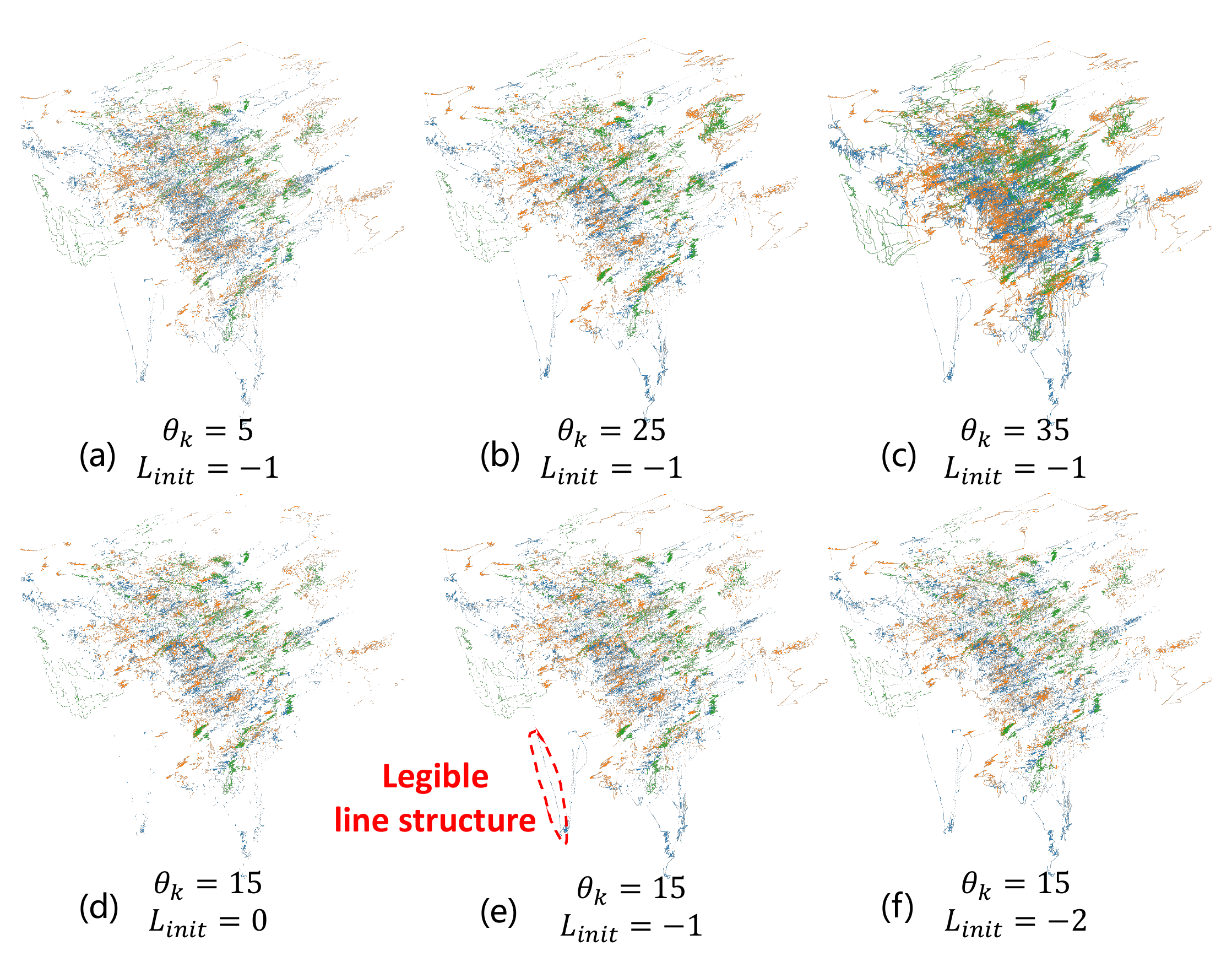}
    \caption{Parameters analyze on \textit{HT sensor}.}
    \label{fig:Parameter Analysis}
    \vspace{-.2in}
    
\end{figure}

The parameter \( \theta_{k} \) influences the granularity of cluster partitioning, as shown in \cref{fig:Parameter Analysis}(a-c, e). For larger \( \theta_{k} \), the cluster partition is more lenient, even if there are some peaks within the region. When the parameter is smaller, we obtain more fine-grained clusters. Thus, smaller \( \theta_{k} \) generally leads to higher contrast. However, too small \( \theta_{k} \) can result in overly fine cluster partitions, causing loss of structural information and increased computational time. Typically, we set \( \theta_{k} \) from 10 to 20, which maintains a good balance between density expression and computation time for most datasets. For sparser datasets or larger canvases, the value should be adjusted downward accordingly.

The parameter \( L_{init} \) helps retain structural information in sparse regions, as shown in \cref{fig:Parameter Analysis}(d-f). When the initial partitioning level is higher (i.e., \( L_{init} \) is set smaller), even low-density regions with dispersed \textcolor{black}{points} can be grouped into one cluster rather than a series of isolated clusters, helping in structural preservation for sparse regions. For example, \cref{fig:Parameter Analysis}(e) retains discrete line structures. However, too small \( L_{init} \) would reduce number of clusters in medium-density regions, which could cause lower contrast. We automate this process based on an empirical baseline of \( L_{init}  = -1\) for a \( 1000\times1000 \)
 canvas. This value is then decreased by 1 for each doubling of the canvas size and increased by 1 for each halving, with the final result rounded to the nearest integer.

\section{Application}

\subsection{High-Definition Display at Any Resolution}

\begin{figure}[tbp]
    \centering
    \includegraphics[width=\columnwidth]{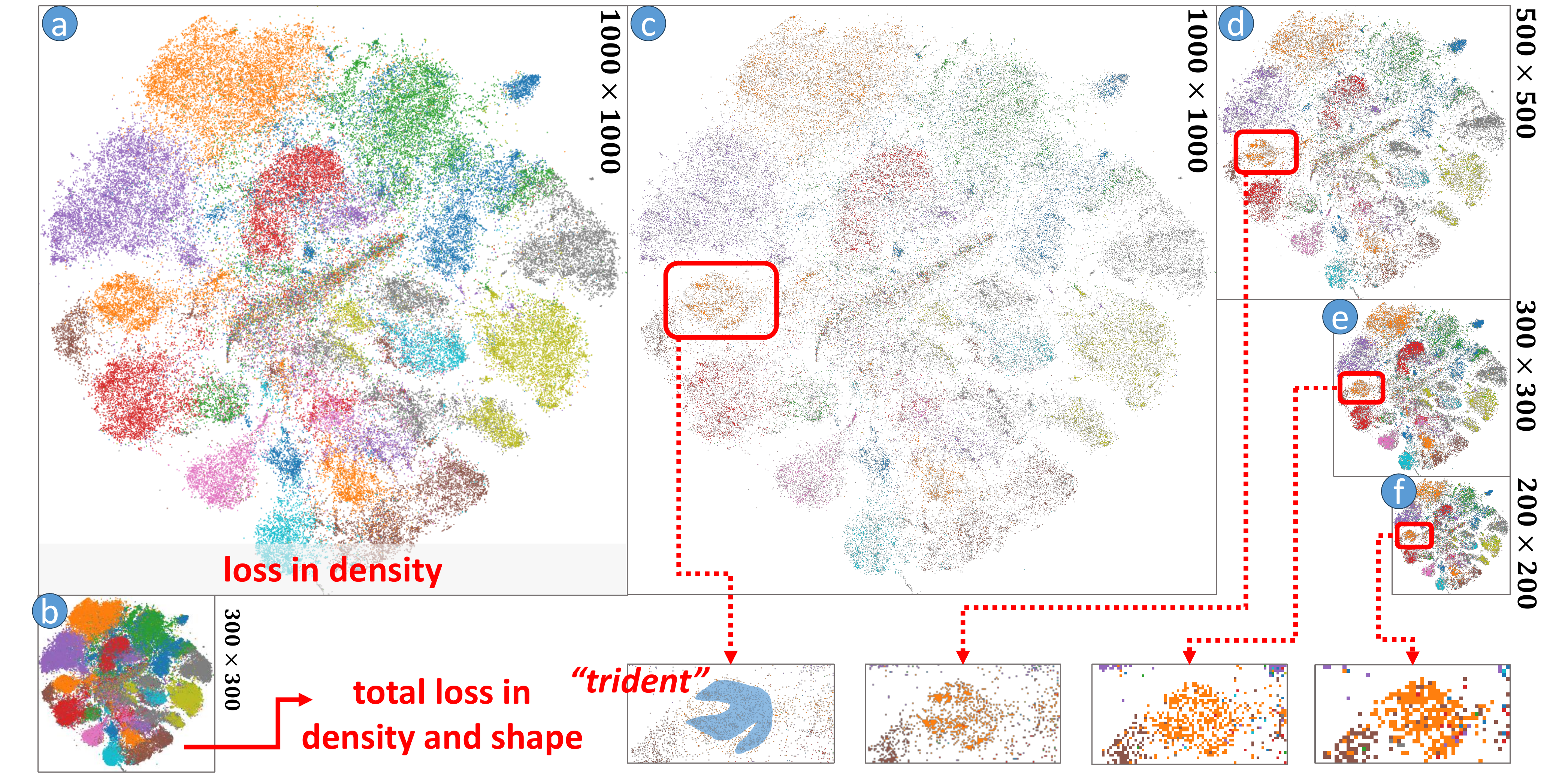}
    \caption{High-definition display at any resolution on $CS ranking$.}
    \label{fig:application -1}
    \vspace{-.2in}
\end{figure}

In scenarios such as printing or covers, there is always a need to provide results at a fixed resolution. Sometimes the resolution may even be extremely low, when in scenarios with network congestion or \textcolor{black}{when thumbnails are required.} Our method can display as much information as possible at any resolution. \cref{fig:application -1}(c-f) shows the results of our method on the $CS ~ ranking$ dataset, with the resolution on the right. For comparison, (a)(b) show the effects of the original scatterplot layout. The most distinct feature of this dataset includes several clusters representing specific disciplines: peripheral clusters are more separated, whereas those near the center are smaller and less distinct. At a normal resolution, the original plot loses density information within the clusters, shown in (a). Our method preserves the density within the clusters, shown in (c). When the resolution reduces as in (b), the original boundaries of the clusters overlap and become indistinguishable. Additionally, colors blend into unrecognizable new colors. Yet, our method maintains the visibility of clusters at the same or lower resolution, and even the internal structure of the clusters is preserved. The orange cluster on the left in (c-f) demonstrates this capability, where we preserve its $"trident"$ shaped structure. Class density and outliers are also maintained, with no color blending.

\subsection{Enhanced Perception of Semantic Outliers}

\begin{figure}[tbp]
    \centering
    \includegraphics[width=\columnwidth]{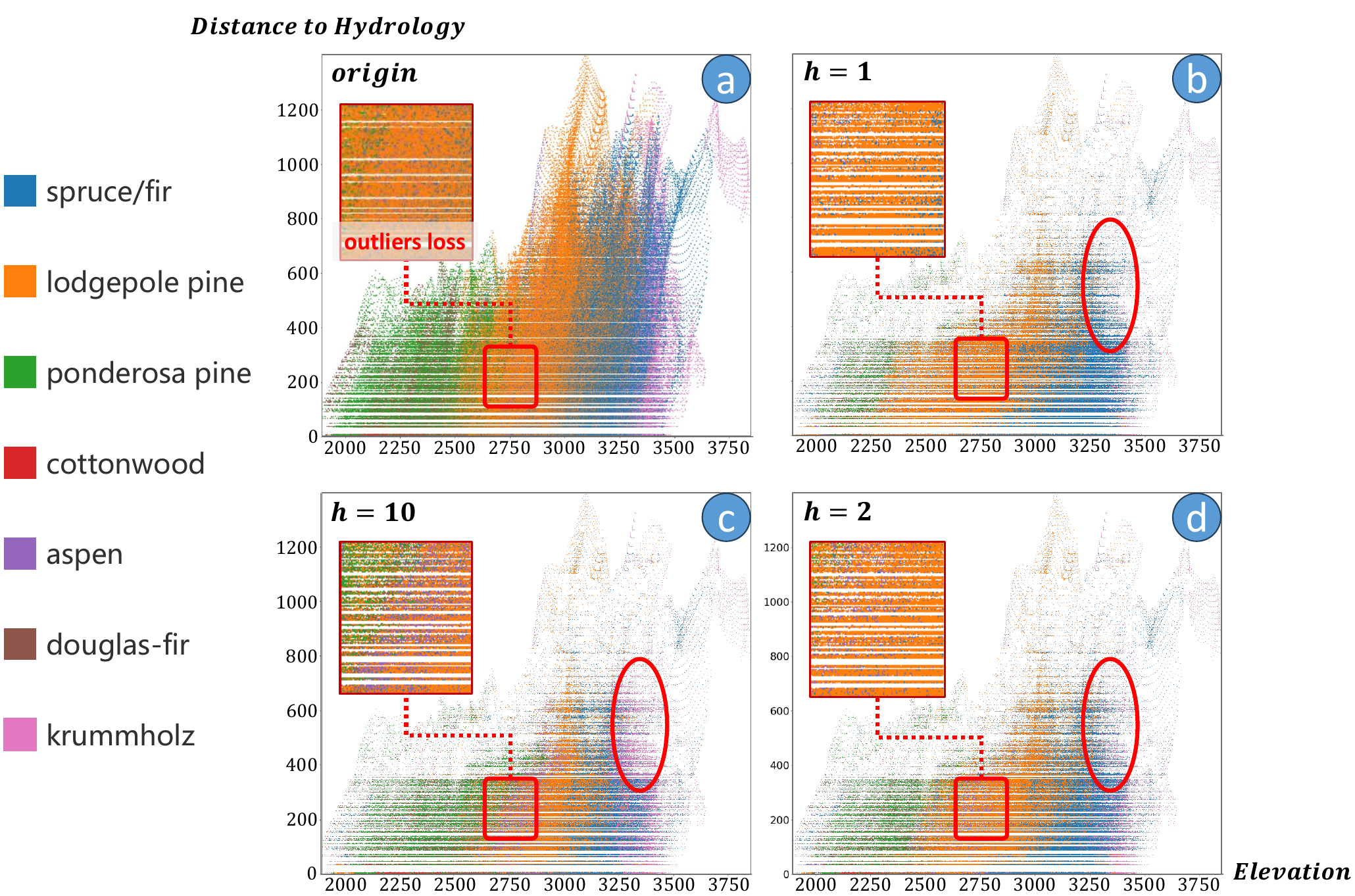}
    \caption{Enhanced perception of semantic \textcolor{black}{outliers} on \textit{Forest covertype}.}
    \label{fig:application -2}
    \vspace{-.2in}
\end{figure}

\textit{Forest covertype} dataset \cite{covertype_31} contains over 581,000 instances. The scatterplot is generated with the x-axis representing \textit{Elevation} and the y-axis representing \textit{Distance to Hydrology}, with forest cover type as the point class, as shown in \cref{fig:application -2}(a). \textcolor{black}{The distribution of forest cover seems to be significantly related to elevation. However, the overlap of points makes it difficult to identify sparse cover types, while the loss of density complicates the perception of coverage rates.}

Through our method, the scatterplot is re-laid out as shown in (b). First, we obtain a clear perception of density: krummholz is primarily distributed at 3400m high, and forest cover is very sparse at this elevation. As the distance to hydrology increases, the forest rapidly diminishes, with the first significant change occurring at about 350m, where the forest cover is nearly halved; by 650m, it further decreases to just a few sparse trees. To better perceive outlier classes, we set $h=2$, as shown in (d). The pink in the circled area, can be identified more clearly. Taking a step further, we set $h=10$, which is nearly the maximum effect achievable in this dataset, as shown in (c). At this setting, the density of outlier classes is close to but does not exceed that of non-outlier classes, and the pink region within the ellipse can be clearly observed. We can observe the boundaries of semantic outlier classes in red box: ponderosa pine mainly spreads at elevations below 2300m, reaching up to about 2800m. Krumholz has a minor distribution from 2650m to 2850m in elevation, starting again in small amounts from 3250m up to the highest elevation.

\section{Discussion and Future Work}

Our method's effectiveness hinges on the perceptibility of its pixel-based output. While pixels offer maximum information granularity, we acknowledge that individual pixels, especially for outliers, can be difficult to discern. Our framework mitigates this through built-in tools like the Outlier Emphasis Pattern and an adjustable abstraction level (i.e., grid size). Beyond these static solutions, perceptibility could be further enhanced through interactive means like magnifying lenses or zooming, which we leave for future work. A related concern is that emphasizing outliers can potentially distort class proportions, creating visual chaos on small canvases. While our ST2 pattern helps prevent this, a more robust future refinement would be to avoid partitioning regions too small to faithfully represent class density.

Building on this work, we have several promising research directions. A seamless hybrid rendering model could be implemented to transition from the proposed pixelated abstraction (for global overview) to traditional point-based scatterplots (for local detail inspection) during zooming. Furthermore, the parallelizable nature of Iso-density Partitioning could be leveraged for significant performance acceleration or streaming data. Finally, the core principle of density equalization could be extended to other visualization variants, such as continuous-variable scatterplots or continuous density fields. For continuous-variable scatterplots, a direct approach is to apply density estimation within each region to generate a local probability density function. New points can then be sampled from this function and placed near original locations to maintain local structure. This sampling can also be tailored to up-sample from tails, thus enhancing outlier perception.

\section{Conclusion}

\textcolor{black}{In this paper, we proposed PixelatedScatter,  a visual abstraction method for large-scale multiclass scatterplots that effectively preserves features across a high dynamic range of data densities. Our method faithfully preserves the global relative regional density relations, enhances visual density contrast, and achieves a good balance between relative class density preservation and outlier highlighting.  Moreover, our method can be suitable for arbitrary abstraction levels. We performed quantitative and qualitative evaluations, as well as a user study. Comparisons with state-of-the-art techniques demonstrate the advantages of our method.}

\acknowledgments{%
The authors wish to thank anonymous reviewers. This work was supported in part by a grant from National Natural Science Foundation of China (\# 62172295).%
}

\bibliographystyle{abbrv-doi-hyperref}

\bibliography{template}

\appendix 

\end{document}